\documentclass[%
 reprint,
superscriptaddress,
 amsmath,amssymb,
 aps,
showkeys,
showpacs
]{revtex4-2}

\usepackage{graphicx}
\usepackage{dcolumn}
\usepackage{bm}

\usepackage{subfiles}
\usepackage{comment}
\usepackage{lipsum}
\usepackage{amsmath}

\begin{document}


\title{Constraints on scalar field dark matter from colocated Michelson interferometers}

\author{Lorenzo Aiello}
\email{Correspondence email address: AielloL@cardiff.ac.uk}
\affiliation{Gravity Exploration Institute, Cardiff University, Cardiff CF24 3AA, United Kingdom}

\author{Jonathan W. Richardson}
\affiliation{Department of Physics and Astronomy, University of California, Riverside, Riverside, California 92521, USA}

\author{Sander M. Vermeulen}
\affiliation{Gravity Exploration Institute, Cardiff University, Cardiff CF24 3AA, United Kingdom}
 
\author{Hartmut Grote}
\affiliation{Gravity Exploration Institute, Cardiff University, Cardiff CF24 3AA, United Kingdom}

\author{Craig Hogan}
\affiliation{University of Chicago, Chicago, Illinois 60637, USA}
\affiliation{Fermi National Accelerator Laboratory, Batavia, Illinois 60510, USA}
 
\author{Ohkyung Kwon}%
\affiliation{University of Chicago, Chicago, Illinois 60637, USA}
 
\author{Chris Stoughton}%
\affiliation{Fermi National Accelerator Laboratory, Batavia, Illinois 60510, USA}

\date{\today}

\begin{abstract}
Low-mass (sub-eV) scalar field dark matter may induce apparent oscillations of fundamental constants, resulting in corresponding oscillations of the size and the index of refraction of solids. Laser interferometers are highly sensitive to changes in the size and index of refraction of the main beamsplitter. Using cross-correlated data of the Fermilab Holometer instrument, which consists of twin co-located 40-m arm length power-recycled interferometers, 
we investigate the possible existence of scalar field dark matter candidates in the mass range between 1.6$\cdot$10$^{-12}$ eV and 1.0$\cdot$10$^{-7}$ eV. We set new upper limits for the coupling parameters of scalar field dark matter, improving on limits from previous direct searches by up to three orders of magnitude. 
\end{abstract}


\maketitle

\section{Introduction}
Cosmological modelling suggests `dark matter' (DM) comprises a large fraction of the total matter content of the universe; its origin and physical properties remain unknown to date. Although indirect astronomical observations in the last decades seem to confirm its existence, a first direct detection of dark matter is still missing and represents one of the greatest challenges of contemporary physics. 

In scenarios where a low-mass scalar field constitutes DM, the field can couple to interferometric detectors, which then produces a signal. This prediction can be used to search for DM signals, and if no signal is detected, constraints on the DM parameters can be placed. Recent examples of this new type of DM search have involved gravitational wave detectors: GEO\,600 \cite{vermeulen2021direct} for scalar field DM \cite{Stadnik_2015,Grote_2019} and Advanced LIGO and Advanced VIRGO \cite{Guo_2019,theligoscientificcollaboration2021constraints} for dark photon DM \cite{Pierce_2018}. Thanks to technological progress, these precision interferometers can now reach sensitivities to length variations at or beyond quantum limits, and they can be used in contexts different from the ones for which they were originally developed.

In this work, we perform a direct search for scalar field DM using the data of the Fermilab Holometer instrument \cite{Chou_2016,Chou_2017_2}, which consists of twin Michelson interferometers. Its cross-correlation sensitivity in the 1-25~MHz frequency range allows us to set new upper limits on the coupling parameters of scalar field DM in a different DM mass range from the one already constrained by the GEO\,600 interferometer and other experiments. 

\section{Theory}
The work presented in this paper concerns low-mass ($m_{\phi}\ll$ 1 eV) DM models. In this scenario, the DM is represented by a scalar field $\phi$ of mass $m_{\phi}$ that couples with the Standard Model (SM) fields. The coupling is parameterised by the addition of an interaction term in the SM Lagrangian \cite{Stadnik_2015_2, Grote_2019}. \\ In this paper we only consider interactions linear in $\phi$ \cite{Grote_2019, vermeulen2021direct}; the resulting Lagrangian is of the form
\begin{equation}\label{eq:lagrangian}
    \mathcal{L}_{\mathrm{int}}^{\mathrm{lin}}=\frac{\phi}{\Lambda_{\gamma}}\frac{F_{\mu\nu}F^{\mu\nu}}{4}-\frac{\phi}{\Lambda_e}m_e\bar{\psi}_e\psi_e
\end{equation}
where $F_{\mu\nu}F^{\mu\nu}$ is the electromagnetic field tensor, $m_e$ the electron rest mass, $\psi_e$, $\bar{\psi}_e$ the SM electron field and its Dirac conjugate. $\Lambda_{\gamma}$ and $\Lambda_e$ parameterise the coupling of the DM field with photons and electrons, respectively. Some scalar field DM models motivated by string theory, such as the Modulus and Dilaton fields, couple to the QCD sector of the SM as well \cite{Damour_1994,Arvanitaki_2016}. These other couplings make such undiscovered massive scalars subject to additional experimental constraints, although the predicted phenomenology due to the electromagnetic coupling terms (Eq.~\ref{eq:lagrangian}) remains unchanged. Relaxions are non-dilatonic scalars which couple to the standard model by mixing with the Higgs field \cite{Flacke_2017}; however, their coupling to the electromagnetic sector can be effectively described by Eq.~\ref{eq:lagrangian} and thus Relaxion DM would produce the same variation of the fundamental constants. \\ 
For sub-eV masses, the field manifests as an oscillating classical field \cite{Stadnik_2015}
\begin{equation}\label{eq:DMfield}
    \phi(t,\vec{r})=\phi_0\mbox{cos}(\omega_{\phi}t-\vec{k}_{\phi}\cdot\vec{r})
\end{equation}
where $\omega_{\phi}=\frac{m_{\phi}c^2}{\hbar}$ is the angular Compton frequency, $\vec{k}_{\phi}=\frac{m_{\phi}\vec{v}_{\mathrm{obs}}}{\hbar}$ the wave vector and $\vec{v}_{\mathrm{obs}}$ is the velocity relative to the observer. Under the assumption that the field is DM, its amplitude is given by $\phi_0=\frac{\hbar\sqrt{2\rho_{\mathrm{local}}}}{m_{\phi}c}$ \cite{Read_2014}, where $\rho_{\mathrm{local}}$ is the local DM density. 

The presence of an oscillating DM field interacting with the SM fields makes the fundamental constants oscillate at the Compton frequency of the DM field. This causes the electron rest mass $m_e$ and the fine structure constant $\alpha$ to be altered as \cite{Stadnik_2015, Stadnik_2015_2, Grote_2019}
\begin{equation}
    m_e \rightarrow m_e\left(1+\frac{\phi}{\Lambda_e}\right), \hspace{1 em} \alpha \rightarrow \alpha\left(1+\frac{\phi}{\Lambda_{\gamma}}\right).  
\end{equation}

By looking for changes of the size $l$ and the refractive index $n$ of a solid \cite{Grote_2019} caused by the oscillatory variation of $\alpha$ and $m_e$, the coupling of DM to SM fields can be probed. The response of a solid to perturbations with a particular driving frequency $\omega$ can be modelled as a driven harmonic oscillator. The amplitude of the size change of the solid is given to first order by:
\begin{equation}
    \frac{\delta l}{l}=\left(-\frac{\delta \alpha}{\alpha}-\frac{\delta m_e}{m_e}\right)\left(1-\frac{\omega^2}{\omega_0^2}\right)^{-1}
\end{equation}
where $\omega_0$ is the angular frequency of the fundamental longitudinal vibrational mode of the solid driven by the scalar field, and we consider a strongly underdamped system. 

The variation of the refractive index of a solid due to the interaction with the DM field is given by \cite{Grote_2019}
\begin{equation}
    \frac{\delta n}{n}\approx -5\cdot 10^{-3}\left(2\frac{\delta \alpha}{\alpha}+\frac{\delta m_e}{m_e}\right).
\end{equation}

A very suitable experimental setup to investigate the presence of changes in the size and refractive index of solids is a laser interferometer. In this instrument a dedicated optic, the beamsplitter, splits the incoming laser beam into the two arms. The beamsplitter has a partially reflective front surface (typically 50$\%$) and an anti-reflective back surface. The light entering one arm is therefore reflected off the front surface, whereas light entering the other arm is transmitted through the beamsplitter. This asymmetry causes changes in the size and refractive index of the beamsplitter to generate a difference between the optical path lengths of the arms $L_x$ and $L_y$. Given the geometry of a Michelson interferometer, the resulting optical path length difference is expected to be given by \cite{vermeulen2021direct}
\begin{equation}\label{eq:OPLdiff}
    \delta(L_x - L_y)\approx \sqrt{2}\left[\left(n-\frac{1}{2}\right)\delta l + l\delta n\right].
\end{equation}



For the current analysis, we use the data from the Fermilab Holometer experiment, which has been constructed to search for exotic quantum space-time correlations \cite{Chou_2017} (see "The Holometer experiment" below).
As the spatial coherence length of light scalar field dark matter ($\lambda_\phi=1/k_\phi$ see Eq.~\ref{eq:DMfield}) is much greater than the separation of the interferometers, possible DM signals in the two intstruments would appear in phase at any time. As dominant sources of noise (i.e. photon shot noise) are incoherent between the two systems, we can take a coherent average of the cross-spectrum over time to increase the signal-to-noise ratio for potential DM signals, which then increases with the square root of the total measurement time. This yields a sensitivity orders of magnitude greater than would be obtained with a single instrument or using only the auto-spectrum of either interferometer. Specifically, the coherently averaged cross-spectral sensitivity lies five orders of magnitude below the single-instrument noise floor for the current data set.      

The magnitude of the expected signal due to scalar field dark matter in the cross-spectrum is then given by considering the optical response of the Holometer to the signal in Eq.~\ref{eq:OPLdiff}, which gives
\begin{widetext}
\begin{equation}
\begin{split}
\delta(L_x-L_y)\approx &\left(\frac{2c l\sqrt{\rho_{\mathrm{local}}}}{\omega_{\phi}}\right)
\left[\mathrm{sinc}\left(\frac{\omega_{\phi} L}{c}\right)\right]^{-1}\left\{\left(n-\frac{1}{2}\right)\left(1-\frac{\omega_{\phi}^2}{\omega_0^2}\right)^{-1}\left(\frac{1}{\Lambda_{\gamma}}+\frac{1}{\Lambda_e}\right)+n\left(\frac{10^{-2}}{\Lambda_{\gamma}}+\frac{5\cdot10^{-3}}{\Lambda_e} \right)\right\},
\end{split}
\label{signal}
\end{equation}
\end{widetext}
where the sinc function describes the modulation of the signal due to the periodic frequency response of an interferometer with arms of length $L$. $\omega_{0}$ is the fundamental angular vibrational frequency of the beamsplitter, which is $2\pi\cdot226$~kHz for the Holometer \cite{Chou_2017}.\\

The mechanical mounting of the beamsplitter has a very minor effect on this resonance. A naive model of the vibrational modes of a simple cylinder, with dimensions and material matching the beamsplitter, predicts a fundamental planar mode frequency of 225 kHz, whereas the measured value of the resonance is 226 kHz; i.e. the mount structure causes a frequency shift of less than 0.5$\%$. A detailed description of this effect is reported in Section 6.5.1 of \cite{Chou_2017}. 

\section{Data analysis and results}



%


The Holometer consists of two independent power-recycled 40~m arm-length Michelson interferometers, coaligned and separated by 0.9 m beamsplitter-to-beamsplitter. In each interferometer, continuous-wave 1064 nm laser light is injected to a beamsplitter and routed along two distinct paths through the interferometer arms to distant end mirrors. 
The instrument is an overcoupled Fabry-Perot cavity whose free spectral range of 3.85 MHz is determined by the average arm length. 
A typical resonating power of 1.3 kW from 1 W of injected power is achieved. A separate radiofrequency data acquisition system samples the interferometer output intensities at 50 MHz and computes their cross-spectral density with 25 MHz bandwidth. A detailed description of the Holometer detector design is provided in \cite{Chou_2017}.

We performed the analysis on a 704-hr dataset acquired between July 2015 and February 2016 \cite{Chou_2016,Chou_2017_2}. During the data taking, the photodetector signals of the two interferometers were sampled at 50 MHz. The time series data was segmented and windowed (overlapping successive segments by 50\% to preserve information \cite{Welch_1967}), after which Discrete Fourier Transforms (DFTs) were computed. Each DFT batch was used to compute a cross-spectrum. Finally, a coherent average over time of all cross-spectra was taken in order to reduce the noise \cite{Chou_2017,Chou_2017_2}. A detailed description of the data acquisition system can be found in Section 5.3 of \cite{Chou_2017}.




We then searched for significant peaks relative to the background noise in the cross-spectrum magnitude. A peak was considered a possible candidate when there was less than 5$\%$ probability that it was due to noise. This probability was determined under the assumption that the noise was Rayleigh distributed and stationary, which has been verified in previous work \footnote{The CSD magnitude is a Rayleigh-distributed quantity, as the real and imaginary parts of the CSD, taken individually, are Gaussian-distributed. The Gaussianity of the Holometer data has been previously established, see Figure 2 of \cite{Martinez_2020} and, in particular, Figure 16 of \cite{Chou_2017}}. The median of the local noise distribution of the time-averaged cross-spectrum was estimated at each frequency bin, using a moving average over neighbouring frequency bins. Different values for the number of neighbouring bins in the moving window (N) were used for different frequency regions \footnote{The choice of N represents a trade-off between erroneously assuming instrumental spectral artefacts or signals to be features of the underlying noise spectrum versus erroneously assuming features of the underlying noise spectrum to be instrumental spectral artefacts or signals. In other words, N has to be chosen such that the auto-correlation length of the noise spectrum is much greater than N. We used different window sizes N in five frequency regions, i.e. N=500 (840 kHz - 25 MHz), N=250  (650 - 840 kHz), N= 70 (250 - 650 kHz), N=20 (14 - 250 kHz) and N=4 (381 Hz - 14 kHz).}

\begin{figure*}[t]  
    \centering
    \includegraphics[width=\textwidth,height=8.8 cm]{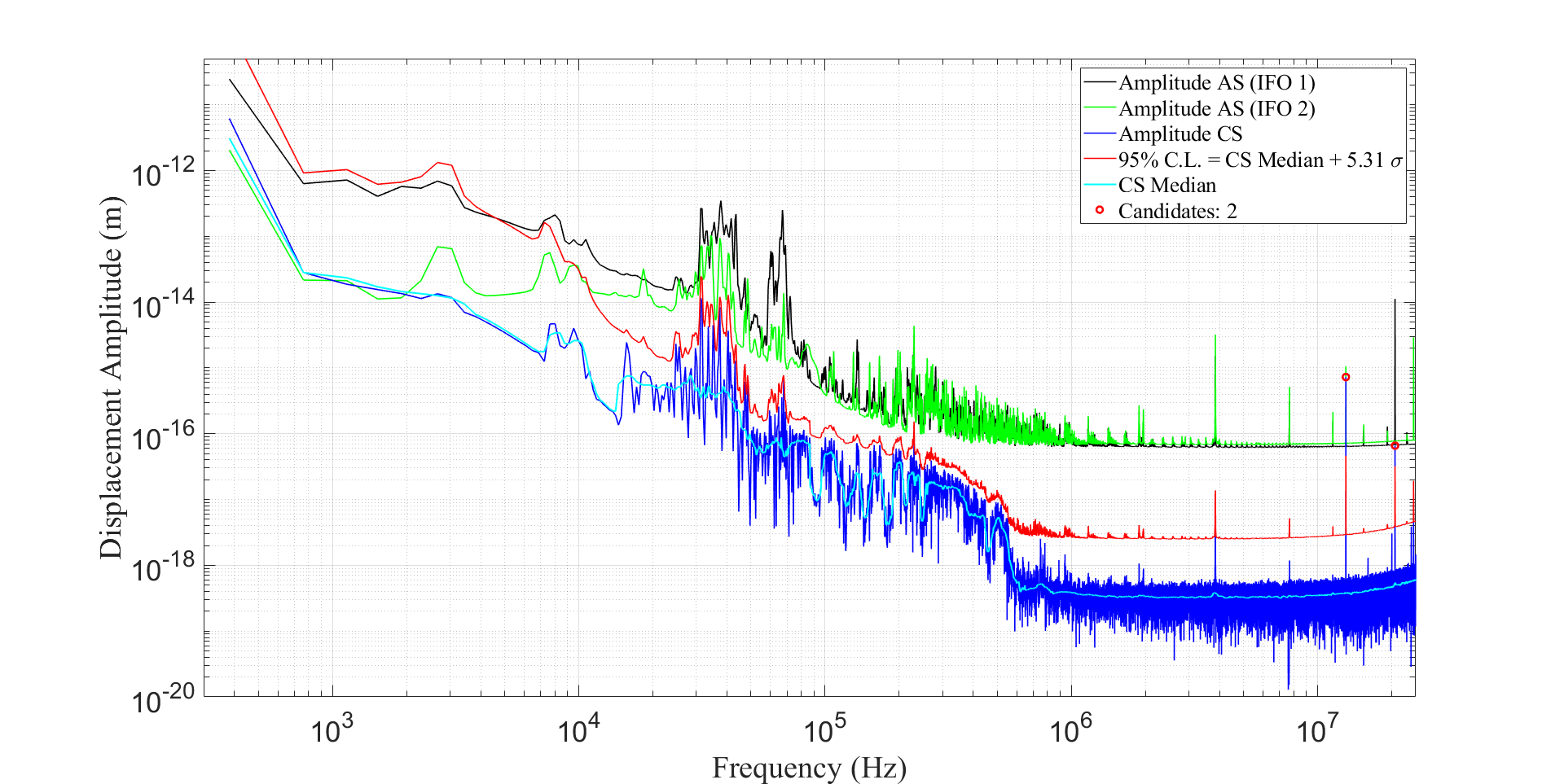}
    \caption{The amplitude auto-spectrum of the single interferometers (black and green) and the cross-spectrum magnitude (blue) obtained from the Holometer's coherently averaged data. At frequencies above $\approx1$~MHz the spectrum is dominated by photon shot noise; below $\approx1$~MHz environmental (mechanical) and laser noise dominate \cite{Chou_2017}. In particular, below 500 kHz the dominant noises are laser amplitude and phase noise - for a detailed description of their characterisation, see Sections 6.4.1 and 6.4.2 of \cite{Chou_2017}. For each frequency bin, the local noise median (cyan) was estimated from its neighbouring bins. The 95$\%$ confidence level (red) was then computed assuming the noise to be Rayleigh distributed. Peaks (red) above the 95$\%$ confidence level were considered possible DM candidates and were investigated further.}
    \label{AS}
\end{figure*}

The frequency-dependent noise variance was estimated directly as the sample variance of all DFTs taken over time. The total error $\sigma$ also includes a calibration error inherent to the apparatus. The look-elsewhere effect was compensated with the application of a trial factor of approximately~$\sim6\cdot10^4$ to account for the number of bins in the cross spectrum. The performed analysis resulted in the identification of two possible candidates above the 95$\%$ confidence level, i.e. $>5.31\sigma$, as shown in Figure \ref{AS}. 

These two peaks were then subjected to further analysis to investigate if either was a DM signal. Both the identified relevant peaks in the amplitude spectrum were related to known harmonic sources inherent to the experiment. The first, at $\sim$13 MHz, is injected for diagnostic monitoring of the readout system by a LED placed directly in front of the photodetectors. The second peak, at $\sim$ 20.5 MHz, is the RF control sideband used to phase lock the lasers to the resonant interferometer cavities \cite{Martinez_2020}. 

Having ruled out the presence of signals due to scalar field DM, we set new constraints on the DM parameters at the 95$\%$ confidence level using Eq. \ref{signal}, applying our analysis to three different DM scenarios.

\begin{figure*}
    \centering
    \includegraphics[width=7.5 cm,height=6.3 cm]{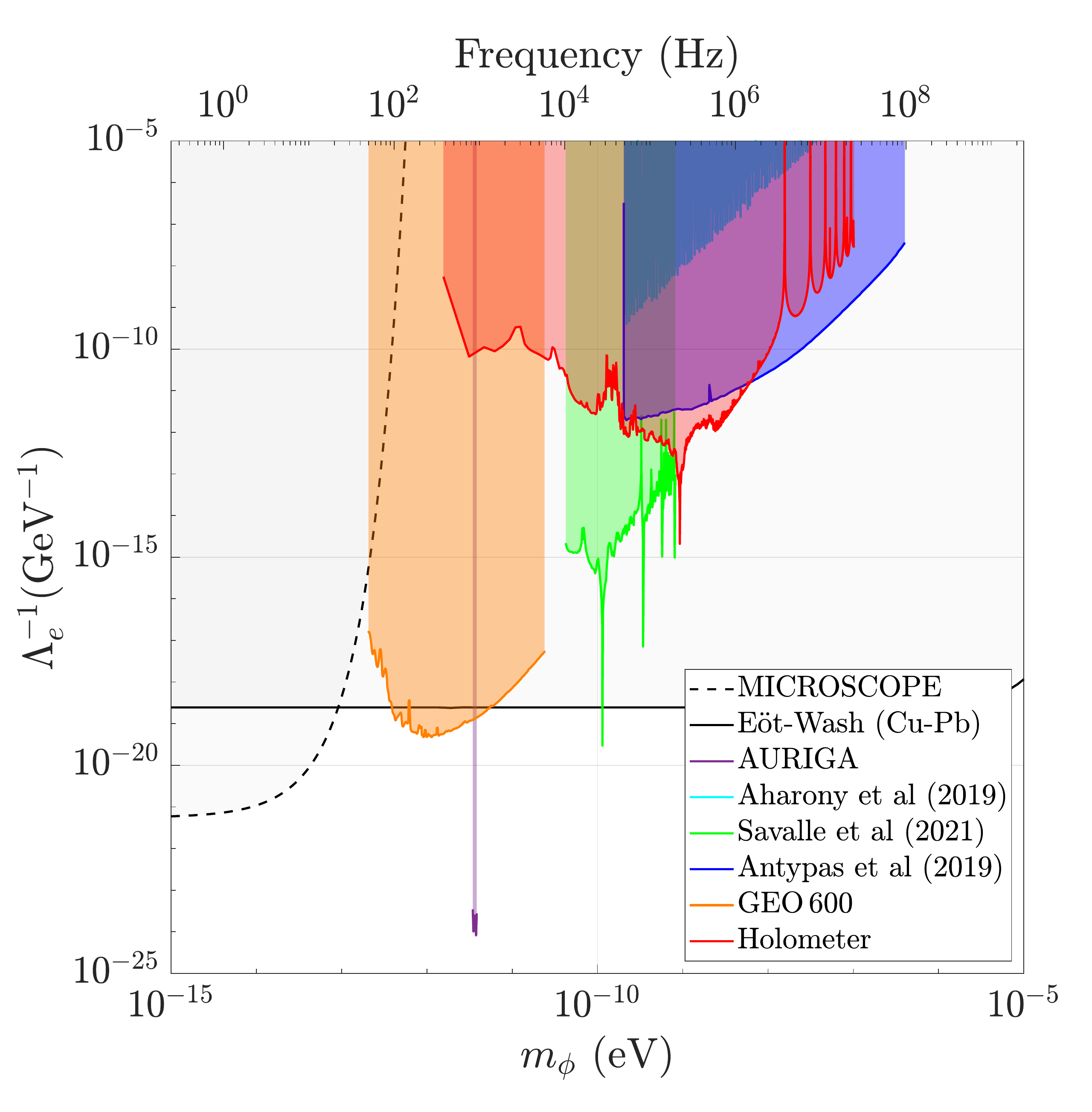}
    \qquad\qquad
    \hspace{-1 cm}
    \includegraphics[width=7.5 cm,height=6.3 cm]{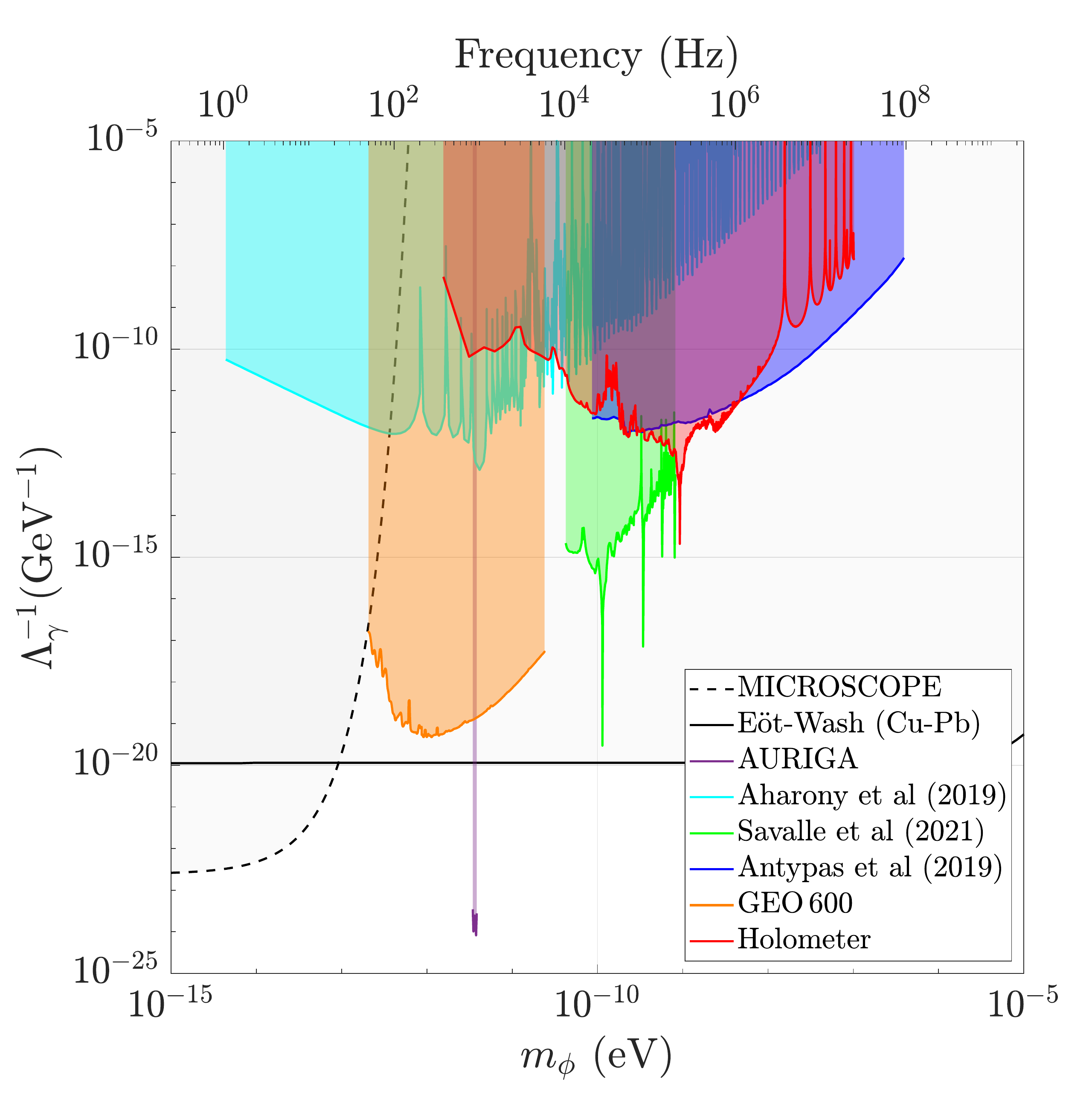}
    \qquad\qquad
    \hspace{-1 cm}
    \includegraphics[width=7.5 cm,height=6.3 cm]{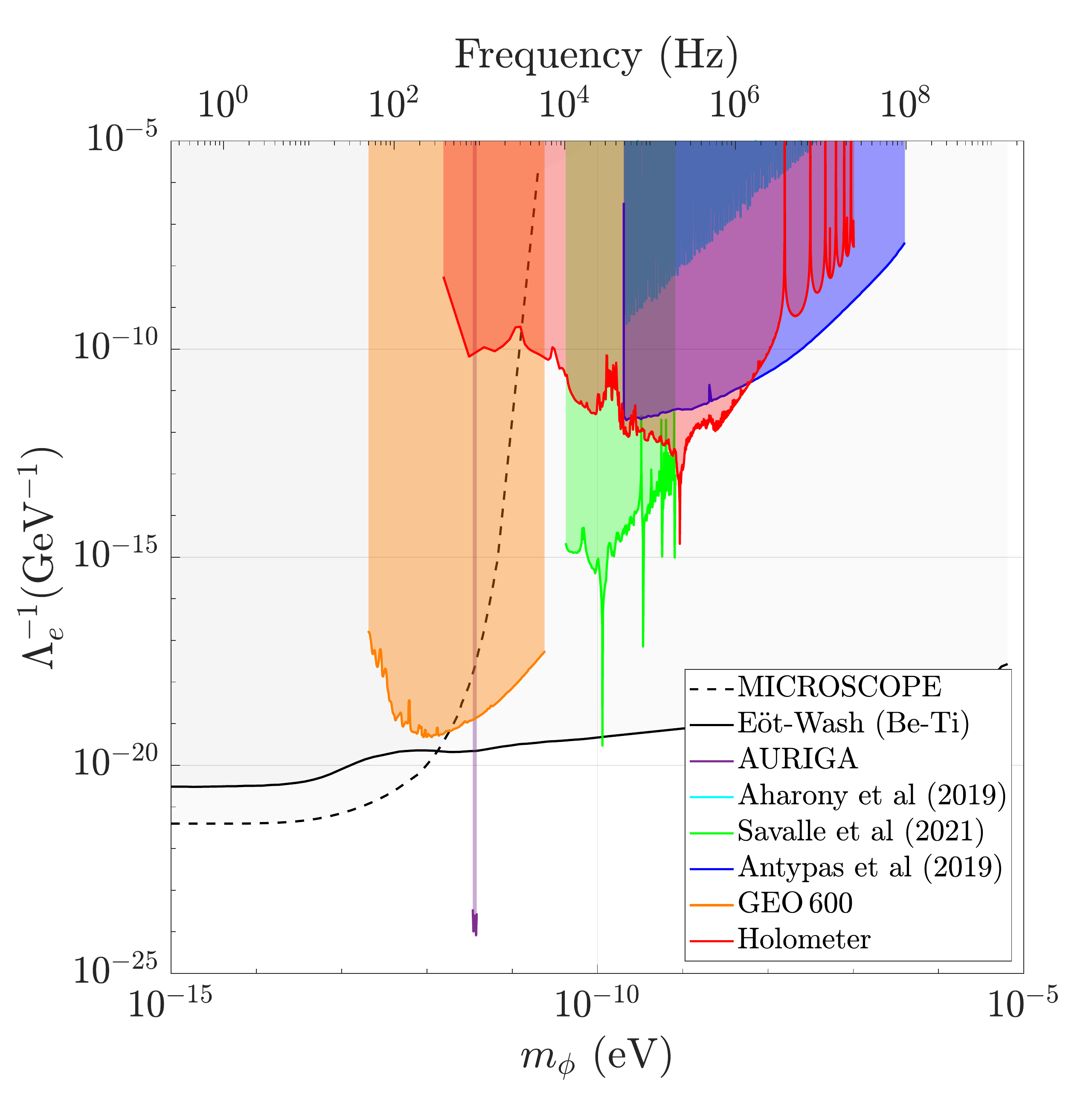}
    \qquad\qquad
    \hspace{-1 cm}
    \includegraphics[width=7.5 cm,height=6.3 cm]{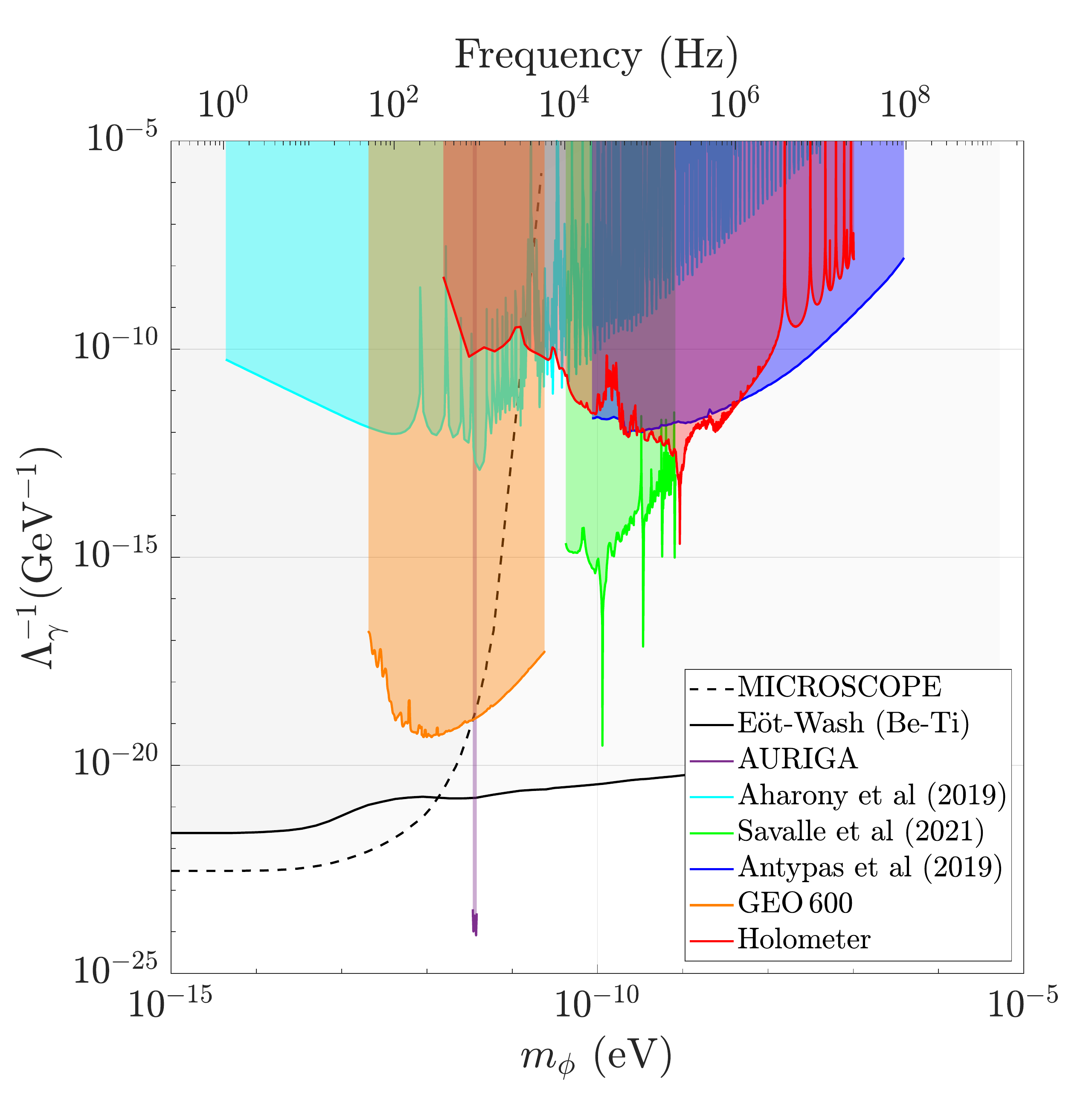}
    \qquad\qquad
    \hspace{-1 cm}
    \includegraphics[width=7.5 cm,height=6.3 cm]{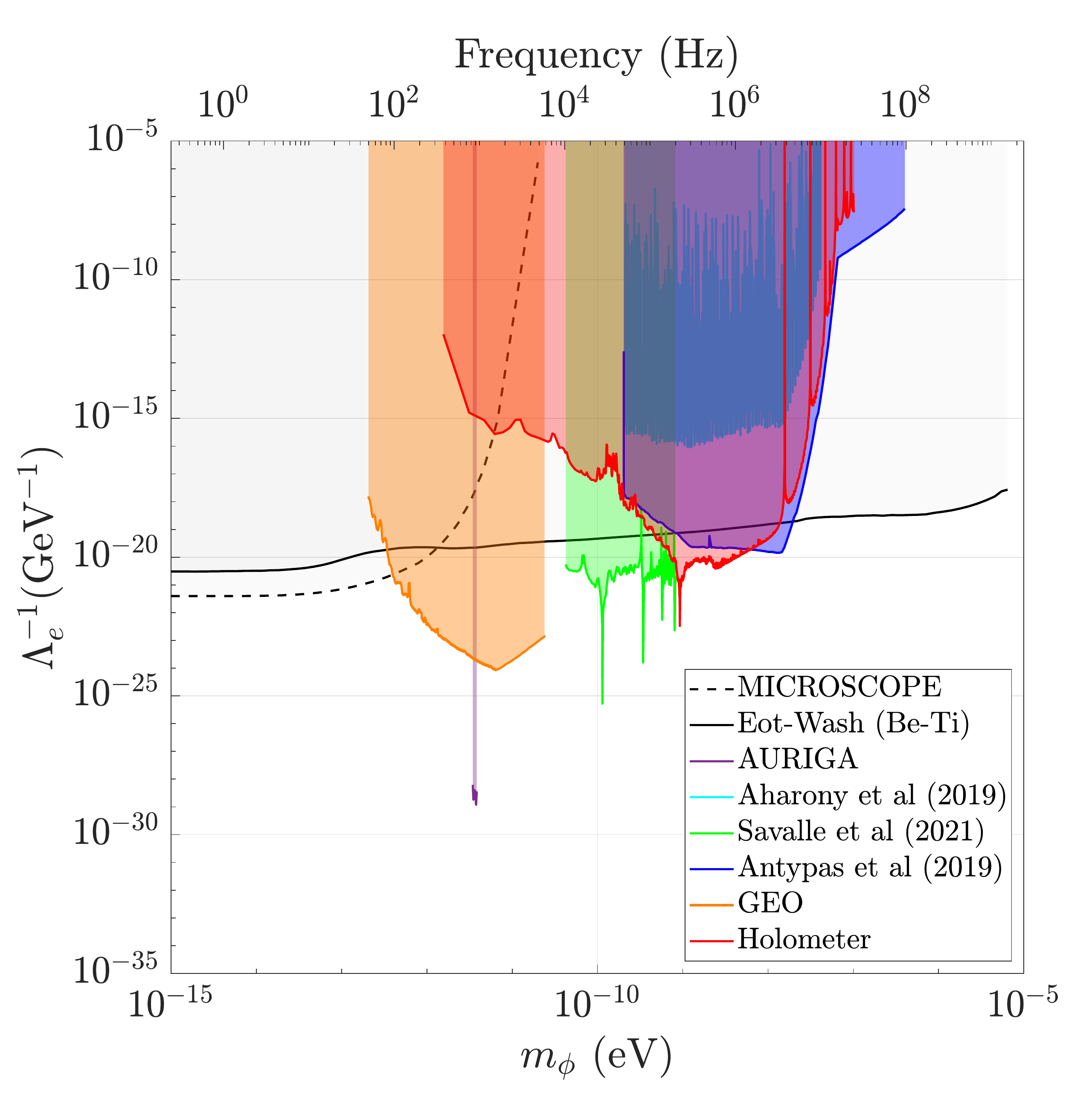}
    \qquad\qquad
    \hspace{-1 cm}
    \includegraphics[width=7.5 cm,height=6.3 cm]{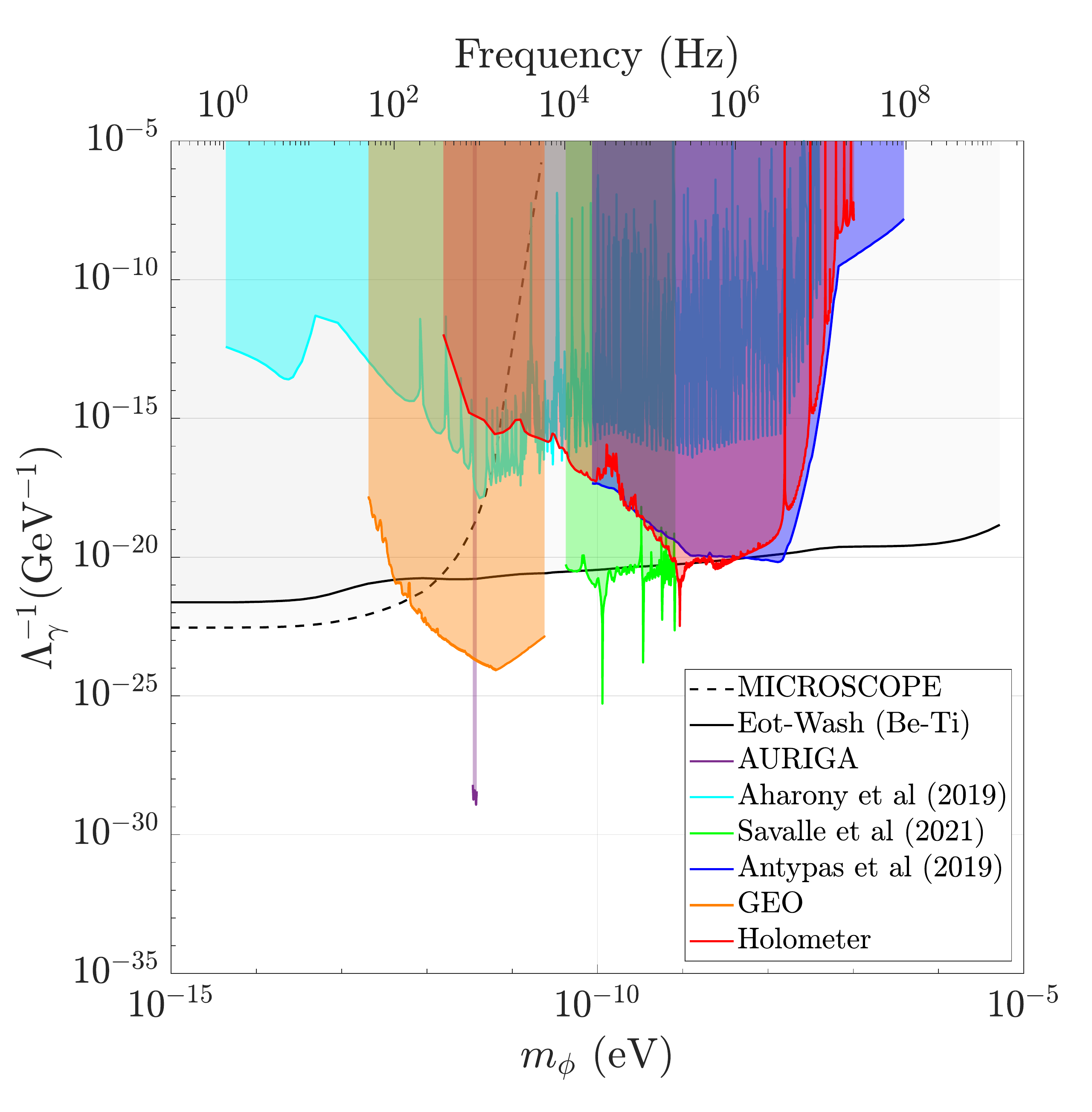}
    \caption{Computed constraints on the coupling parameters $\Lambda_e$ (left) and $\Lambda_{\gamma}$ (right) as a function of the field's mass $m_{\phi}$ for scalar field DM as in the Basic Scalar scenario (top), the Dilaton/Modulus scenario (middle) and the Relaxion Halo scenario (bottom). Electron and photon coupling constraints are at the 95$\%$ confidence level. The region coloured in red indicates the parameter space for the coupling parameters excluded by our analysis of the Holometer data. Other coloured regions mark the parameter space excluded by other direct searches \cite{Aharony_2021,Savalle_2021,Antypas_2019}, including the AURIGA experiment \cite{Branca_2017} and the GEO 600 interferometer \cite{vermeulen2021direct}. The grey regions denoted by the black curves are constraints on general fifth-forces and tests of the equivalence principle \cite{Leefer_2016}. These come from the space-based MICROSCOPE experiment \cite{Berg__2018}, and the Cu/Pb and the Be/Ti torsion pendulum experiments performed by the E$\ddot{\mbox{o}}$t-Wash group \cite{Smith_1999,Wagner_2012,Schlamminger_2008}. For the Relaxion Halo scenario, a mass-dependent DM halo density as described in \cite{Banerjee_2020} has been assumed. The constraints obtained for this scenario from direct experimental searches have been obtained by rescaling the original ones to account for this dependence. Constraints from fifth-force and equivalence principle tests do not depend on the local DM density and are thus the same as in the Dilaton/Modulus scenario.}
    \label{Constraints}
\end{figure*}


The first scenario is the \emph{Basic Scalar}, see Fig. \ref{Constraints} (top). In this scenario the interaction between the scalar field DM and the SM fields is fully described by Eq. \ref{eq:DMfield}. The scalar field is assumed to be homogeneous in the solar system and, according to the standard galactic DM halo model \cite{Read_2014}, its density is equal to $\rho$=0.4 GeV/cm$^3$.

The second scenario is the \emph{Dilaton/Modulus}, see Fig. \ref{Constraints} (middle). The density of the DM field is assumed to be the same as in the Basic Scalar model. 
With respect to the previous scenario, the Dilaton/Modulus is constrained by additional limits from Equivalence Principle tests, but is equally constrained by our results. 

The third scenario is the \emph{Relaxion Halo}, see Fig. \ref{Constraints} (bottom). Here it is assumed that a Relaxion Halo gravitationally bound to Earth dominates the local DM density, and leads to a local overdensity (which depends on the field’s mass) that can reach values of up to $\frac{\rho_{local}}{\rho}\sim$10$^{19}$ for the mass range constrained by our analysis \cite{Banerjee_2020}. The coupling between the DM scalar field and the SM field is as in the Dilaton/Modulus scenario (although these couplings arise through mixing with the Higgs boson).

The electron and photon coupling parameters, $\Lambda_e$ and $\Lambda_{\gamma}$, respectively, are constrained for each scenario as a function of the field's mass $m_{\phi}$, assuming for each coupling parameter the other to be zero. The new constraints obtained from our analysis, together with previously published upper limits, are plotted in Fig. \ref{Constraints}. The feature at 226~kHz is due to the mechanical resonance of the beamsplitter, where the apparent depth of the minimum is limited by the frequency resolution (the Q-factor of the beamsplitter is more than an order of magnitude greater than the plotted amplitude enhancement).

\section{Conclusions}
In this paper we have looked for signals of scalar field DM in the cross-spectrum of 
co-located interferometers, which constitutes the first direct search of scalar DM using correlated interferometry.  


Our analysis excluded the presence of scalar field DM signals in the data, placing lower limits on the DM coupling parameters 
for DM masses between 1.6$\cdot$10$^{-12}$ eV and 1.0$\cdot$10$^{-7}$ eV. These limits improve over previous direct experimental bounds in several subranges: we set limits in the previously unconstrained mass range between $2.4\cdot10^{-11}$ and  $4.3\cdot10^{-11}$~eV, and improve over the existing constraints \cite{Antypas_2019} in the mass range $8.2\cdot10^{-9}-6.2\cdot10^{-8}$~eV by up to three orders of magnitude. 

Constraints on the coupling parameters of scalar field DM have been previously obtained in many distinct mass ranges with different types of experiments: optical cavities \cite{Savalle_2021}, atomic spectroscopy \cite{Aharony_2021,Antypas_2019}, and GW detectors like the resonant bar AURIGA \cite{Branca_2017} and the GEO\,600 interferometer \cite{vermeulen2021direct}. Finally, equivalence principle tests exploiting torsion balances \cite{Smith_1999,Wagner_2012} have been used to compute constraints on scalar fields under the assumption they manifest as a `fifth force' and are sourced by a test mass \cite{Berg__2018,Hees_2018}. These constraints are independent of the local DM density, and depend in general on the composition and geometry of the test masses.

Better constraints on scalar field DM can be achieved
through upgrades of current experiments \cite{Antypas_2019}, by increasing the measurement time of correlated instruments, or from new, more sensitive experiments (e.g. \cite{Vermeulen_2021}).


\begin{acknowledgments}
L.A. and S.M.V. thank Aldo Ejlli for useful discussions and comments on this work. The authors thank Lee McCuller for valuable comments on the manuscript, and Stephan Meyer for setting up a machine to host the data for this analysis. We thank the Science and Technology Facilities council in the UK for support under grants ST/V00154X/1 and  ST/T006331/1, and the Leverhulme Trust in the UK for support under grant RPG-2019-022. This work was supported by the Fermi National Accelerator Laboratory (Fermilab), a U.S. Department of Energy, Office of Science, HEP User Facility, managed by Fermi Research Alliance, LLC (FRA), acting under Contract No. DE-AC02-07CH11359. We are also grateful for support from the John Templeton Foundation.
\end{acknowledgments}

\bibliography{Biblio}
\bibliographystyle{apsrev4-2}

\end{document}